\documentclass[12pt,letterpaper]{article}

\usepackage[utf8]{inputenc}
\usepackage[T1]{fontenc}
\usepackage{amsmath, amssymb, amsfonts}
\usepackage{graphicx}
\usepackage{float}
\usepackage{setspace}
\usepackage{enumitem}
\usepackage{authblk}
\usepackage{parskip}
\usepackage{xcolor}
\usepackage{siunitx}
\usepackage{hyperref}
\usepackage{geometry}
\usepackage{natbib}
\usepackage{booktabs}
\usepackage{caption}
\usepackage{subcaption}
\definecolor{linkcolor}{RGB}{0, 90, 160}
\hypersetup{
    colorlinks=true,
    linkcolor=linkcolor,
    urlcolor=linkcolor,
    citecolor=linkcolor
}

\title{Beyond Expected Goals: A Probabilistic Framework for Shot Occurrences in Soccer}
\author[1]{Jonathan Pipping-Gam\'on}
\author[2]{Tianshu Feng}
\author[1]{Paul Sabin}
\affil[1]{\small Department of Statistics \& Data Science, University of Pennsylvania}
\affil[2]{\small Department of Computer \& Information Science, University of Pennsylvania}
\date{\today}

\begin{document}

\maketitle

\begin{abstract}
Expected goals (xG) models estimate the probability that a shot results in a goal from its context (e.g., location, pressure), but they operate only on observed shots. We propose xG+, a possession-level framework that first estimates the probability that a shot occurs within the next second and its corresponding xG if it were to occur. We also introduce ways to aggregate this joint probability estimate over the course of a possession. By jointly modeling shot-taking behavior and shot quality, xG+ remedies the conditioning-on-shots limitation of standard xG. We show that this improves predictive accuracy at the team level and produces a more persistent player skill signal than standard xG models.
\end{abstract}

\section{Introduction}

\subsection{Expected Goals}

Expected Goals (xG) has become the most commonly used metric in modern soccer analytics \citep{eggels2016expected}. This statistic can now be seen on television broadcasts and is even shown as part of the popular video game EA Sports FC (formerly FIFA). 

xG quantifies the probability that a shot will result in a goal based on characteristics such as shot location, angle to the goal, shot type, and defensive pressure \citep{spearman2018beyond}. Perhaps the earliest use of expected goals was \cite{ensum2004applications}, which used a logistic regression model to estimate the probability that a shot becomes a goal. \cite{macdonald2012expected} implements an expected goals model in a different low-scoring sport, ice hockey, where expected goals were used to estimate an adjusted plus-minus model for players in the NHL. Like this work in the NHL, xG in soccer has been shown to be a more predictive metric of future goals scored than actual goals scored (\cite{heuer2012perfect} \& \cite{mead2023expected}). 

\cite{lucey2015quality} introduced the use of spatio-temporal data to estimate expected goals models. \cite{fernandez2019decomposing} uses an expected goals model (among others) to decompose the game into a series of decisions and actions by each player.

\subsection{Limitations of Expected Goals}

Any limitations of xG are important to recognize as xG is not only used as a metric itself but is a foundational piece to many important research papers in soccer analytics. \cite{singh_xT_blog}, \cite{bransen2018measuring}, and \cite{statsbomb2021obv} use the xG value of actual shots to value on-ball actions in the buildup. This is often referred to as expected threat (xT) and sometimes as on-ball value (OBV).

Despite its widespread use, xG remains limited by its foundational assumption: it only evaluates the quality of shots that are \emph{actually taken}. As a result, the model ignores many of the most dangerous moments in a match simply because they did not result in a shot.

There have been very few previous attempts to explicitly model the probabilistic nature of shot taking. The work of \cite{fernandez2019decomposing} and \cite{fernandez2021framework} decomposes actions into a sequence using spatio-temporal tracking data and models the probability of a pass, ball drive, or shot at the end of a possession. The authors spend little time talking about the expected goals part of their work, focusing on the passing and ball drive components, but do report that they had some model for the probability a shot occurs.

Another work that incorporated some probabilistic component of shot taking is  \cite{poropudas2021extended}, who incorporated a shot decision model into expected threat (xT).

A few attempts at adjusted plus-minus models in soccer using expected goals have also been attempted, similar to \cite{macdonald2012expected} in hockey. \cite{matano2018augmenting} attempted to use expected goals in their FIFA rating augmented plus-minus but decided against it due to data limitations. \cite{kharrat2020plus} and \cite{zhang2022cmsac} use expected goals to derive various extensions to the augmented plus-minus models proposed by \cite{macdonald2012expected}. 

xG is also typically used in predictive models either alongside or instead of the actual goals in the match, as was the case with FiveThirtyEight's Soccer Power Index (SPI) \citep{fivethirtyeight_clubsoccerpredictions}.

While ``all models are wrong,'' any shortcomings of expected goals models propagate to power ratings models, adjusted plus-minus models, and expected threat models because of their reliance on xG as an outcome variable.

Soccer matches are filled with sequences that nearly result in shots: crosses that are barely intercepted, passes to open attackers that arrive just a moment too late, or dribbles into the box that are stopped by a last-ditch tackle. These moments reflect true offensive danger, yet go unrecorded in traditional xG models. 

Soccer may also fall victim to the same selection bias that plagues expected points models in American football. \cite{brill2025analytics} brought attention to this issue and showed that among other issues, the fact that better offensive teams had more plays closer to the end zone affected the certainty of machine learning-based expected points models and those that relied on expected points (such as 4th down models).

Similarly in soccer, it is possible that players who are better at converting shots to goals take more shots than those who are less skilled. In a similar vein, players who are better at converting opportunities into shots are likely taking an outsized share of shots compared to other players who play similar positions.

Since xG models are trained on recorded shots, any bias in shot takers will also propagate to other models that rely on xG.

Another assumption of xG models is that they treat shots as independent events, so aggregating the metric yields inflated cumulative values whenever a sequence includes multiple rapid-fire rebound attempts, even though only one goal could possibly result per possession.

\subsection{Our Contribution}

To address several of these shortcomings, we propose a new framework that models not just the quality of shots taken, but also the probability of a shot occurring in the first place. This component of the goal-generating process yields the metric \textbf{xS}, which we define as the probability that a shot occurs within the next second. Combining this with the existing xG metric yields \textbf{xG+}, which reflects the probability of scoring in the next second \textbf{whether or not a shot actually occurs}. We also present possession-adjusted xG+, which measures, at each frame, the marginal increase in the probability of scoring by the end of the possession. This adjustment of xG at the possession level recognizes the fundamental limitation that only one goal can be scored per possession.
  
By accounting for both shot generation and goal scoring, xG+ represents a more complete metric that better aligns with how fans, coaches, and analysts intuitively understand the game, while improving prediction of future goal-scoring performance: for players as well as teams.

\section{Motivating Examples}

To illustrate the shortcomings of traditional xG and further motivate our approach, we examine a few concrete match scenarios.

\begin{figure}[H]
\centering
\begin{subfigure}{0.9\textwidth}
  \centering
  \includegraphics[width=\textwidth]{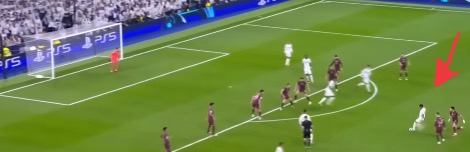}
  \caption{Rodrygo takes a shot from distance with a low probability of scoring.}
  \label{fig:Rodrygo}
\end{subfigure}

\vspace{0.5cm} 

\begin{subfigure}{0.9\textwidth}
  \centering
  \includegraphics[width=\textwidth]{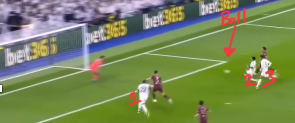}
  \caption{Three players were close to tapping this cross into a goal, but none got to the ball.}
  \label{fig:MissedCross}
\end{subfigure}

\caption{A comparison of a shot with low goal probability and a cross with a much higher goal probability that never became a shot.}
\label{fig:RMChancesCompare}
\end{figure}

On February 19, 2025, Real Madrid faced Manchester City in the second leg of a Champions League knockout round match (Figure \ref{fig:RMChancesCompare}). Early in the second half, Real Madrid's Rodrygo attempted a speculative shot from 35 yards out, which generated a very low xG (approximately 0.03 according to \href{https://fbref.com/en/matches/bef7ecbf/Real-Madrid-Manchester-City-February-19-2025-Champions-League}{FBRef}). Moments later, a cross into the six-yard box nearly connected with a Real forward who was fighting with the defender to get a foot on the ball. If the attacker beats the defender to the ball, it's an almost-sure goal. If the defender gets there first instead, no shot occurs at all. The City defender reached the ball just in time, and no shot was recorded -- resulting in a zero contribution to the team's xG despite clearly being the more dangerous moment.

This contrast illustrates a core flaw in traditional xG: the most threatening moments are not always those that result in shots. Our framework captures this by assigning a nonzero scoring probability to such near-opportunities by accounting for both the probability that a shot occurs and the conditional probability of a goal if a shot occurs, based on features of the tracking data.

\begin{figure}[H]
\centering
\begin{subfigure}{\textwidth}
  \centering
  \includegraphics[width=0.9\textwidth]{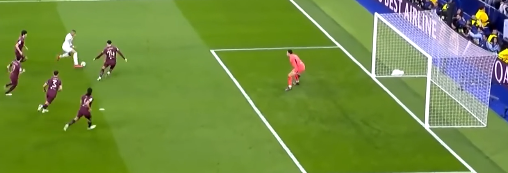}
  \caption{Mbappé has an opportunity to shoot here with a defender bearing down on him.}
  \label{fig:MbappeCreate1}
\end{subfigure}

\vspace{0.5cm} 

\begin{subfigure}{\textwidth}
  \centering
  \includegraphics[width=0.9\textwidth]{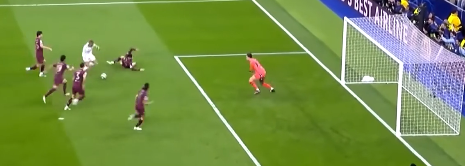}
  \caption{After making his defender miss, Mbappé now has a much better chance to score (0.5 xG according to \href{https://fbref.com/en/matches/bef7ecbf/Real-Madrid-Manchester-City-February-19-2025-Champions-League}{FBRef})}
  \label{fig:MbappeCreate2}
\end{subfigure}

\caption{Kylian Mbappé demonstrates his ability to create high-quality shots by making his man miss against Manchester City (Feb 19, 2025). Traditional xG metrics only consider the probability of a goal once he shoots, which is why some elite goal-scorers fail to consistently outperform their xG. Their xG is high because they created better chances!}
\label{fig:mbappeShotCreate}
\end{figure}


We now present another motivating example that illustrates the need to account for the sequential nature of possessions. In the 78th minute of the February 22, 2025 match between Orlando City and Philadelphia Union, Orlando generated a chaotic attacking sequence that included four shots in rapid succession (per \href{https://fbref.com/en/matches/744e06cd/Orlando-City-Philadelphia-Union-February-22-2025-Major-League-Soccer}{FBRef}). As shown in Table \ref{tab:orlando}, the total xG assigned to this sequence was 1.63, which contradicts the basic logic that an attack can score no more than one goal. This sequence -- featuring blocked shots, rebounds, and ultimately a goal -- highlights the compounding nature of xG in tightly-clustered shot events.

Properly aggregating goal scoring opportunities across a possession is essential to our methods because we evaluate the probability of a goal every second.

\begin{table}[H]
\centering
\caption{Sequence of shots leading to one goal but totaling 1.63 xG.}
\label{tab:orlando}
\begin{tabular}{|c|l|c|}
\hline
\textbf{Time} & \textbf{Player} & \textbf{Shot outcome (xG)} \\
\hline
78:01 & Brekalo & Shot blocked (0.05) \\
78:02 & Muriel  & Shot post (0.52) \\
78:04 & Pasalic & Shot post (0.68) \\
78:05 & Pasalic & Shot goal (0.38) \\
\hline
\end{tabular}
\end{table}

\section{Methodology}

\subsection{Modeling Framework}

To correct for these limitations, we define \texttt{xG+} as the product of two estimated probabilities: the likelihood of a shot occurring (\texttt{xS}) and the likelihood of a goal given that a shot occurs (\texttt{xG}) in the next second. Formally, at any frame $t$:
\[
\text{xG+}_t = \mathbb{P}_t(\text{Shot}) \times \mathbb{P}_t(\text{Goal} | \text{Shot}) = \text{xS}_t \cdot \text{xG}_t
\]

To compute the expected goal probability over an entire possession of $n$ frames, we use the formula:
\[
\text{xG+}_{\text{poss}} = 1 - \prod_{t=1}^{n} \left(1 - \text{xG+}_t\right)
\]

This equation ensures that the total possession value is bounded by one and reflects both latent and realized scoring threats. The focus of this work now turns to estimating the probability of a shot occurring within the next second using the camera-based optical tracking data from Gradient Sports.

\subsection{Data and Feature Engineering}

We use video tracking, event, and team data provided by Gradient Sports for the 2022--23, 2023--24, and 2024--25 English Premier League (EPL) seasons. Event data encodes on-ball actions such as possession changes, shots, and goals, while tracking data includes synchronized ball and player positions.

For each video frame (30 fps), the full dataset includes:
\begin{itemize}
  \item Ball and player positions (x, y, z)
  \item Possession indicators and shot outcomes
  \item Player and team IDs
\end{itemize}

From these raw inputs, we filter our data for sequences where one team has clear possession in the attacking third, then derive the following features:
\begin{itemize}
  \item Ball distance to the goal, bearing (angle) to goal, speed, and height
  \item Relative positioning and Euclidean distances from the ball to the 5 nearest attackers and non-goalkeeper defenders \footnote{Note that the ball should mostly overlap with the attacker closest to the ball since the cleaned dataset is filtered on the condition where one team has clear possession in their attacking third. As a result, the attacker closest to the ball is not counted.}
  \item Goalkeeper location and \texttt{openGoal}, a proxy for how obstructed the path to the goal is.
\end{itemize}

A full description of engineered features is included in Appendix \ref{app:a}.

\subsection{Model Estimation} \label{sec:model-estimation}

We train two separate XGBoost models on our engineered features:
\begin{enumerate}
  \item \textbf{xS}: Predicts the probability that a shot will occur in the next second
  \item \textbf{xG}: Predicts the probability that a shot taken in the current frame results in a goal.
\end{enumerate}

Both models use 5-fold cross-validation within each season, with log loss as the primary evaluation metric. Model results and diagnostics are included in the next section.

\section{Results and Evaluation}

\subsection{Baseline Comparison}

We benchmark our models against logistic regression baselines trained on four feature sets: (i) ball distance only, (ii) all ball features, (iii) ball + goalkeeper features, and (iv) all features. Each model is evaluated with 5-fold cross-validation on identical splits, and we compare mean out-of-sample log loss. As shown in Table \ref{tab:lr_loss}, our XGBoost models outperform all logistic baselines -- especially on the xS task. Our xG model is conceptually similar to the proprietary models described by \citet{hudl_xg_explained}.

\begin{table}[htbp]
\centering
\caption{Model comparison: mean out-of-sample log loss ($\pm$ SD)}
\label{tab:lr_loss}
\begin{tabular}{ll
                S[table-format=1.3]
                S[table-format=1.4]}
\toprule
\textbf{Model} & \textbf{Features} & {\textbf{xG (log loss ↓)}} & {\textbf{xS (log loss ↓)}} \\
\midrule
Logistic  & Ball distance only & \ensuremath{0.356 \pm 0.0105} & \ensuremath{0.0260 \pm 0.00053} \\
Logistic  & Ball (all)         & \ensuremath{0.350 \pm 0.0087} & \ensuremath{0.0260 \pm 0.00053} \\
Logistic  & Ball + GK          & \ensuremath{0.340 \pm 0.0086} & \ensuremath{0.0258 \pm 0.00054} \\
Logistic  & All features       & \ensuremath{0.337 \pm 0.0086} & \ensuremath{0.0251 \pm 0.00060} \\
\addlinespace
\bfseries XGBoost & \bfseries All features & \bfseries \ensuremath{0.326 \pm 0.0074} & \bfseries \ensuremath{0.0227 \pm 0.00089} \\
\midrule
\multicolumn{2}{l}{\emph{Improvement vs best logistic (all features)}} 
  & \multicolumn{1}{c}{\emph{0.011 (3.3\%)}} 
  & \multicolumn{1}{c}{\emph{0.0024 (9.5\%)}} \\
\bottomrule
\end{tabular}
\end{table}

\subsection{Feature Importance and Partial Dependence}

\begin{figure}[H]
  \centering
  \includegraphics[width=0.9\textwidth]{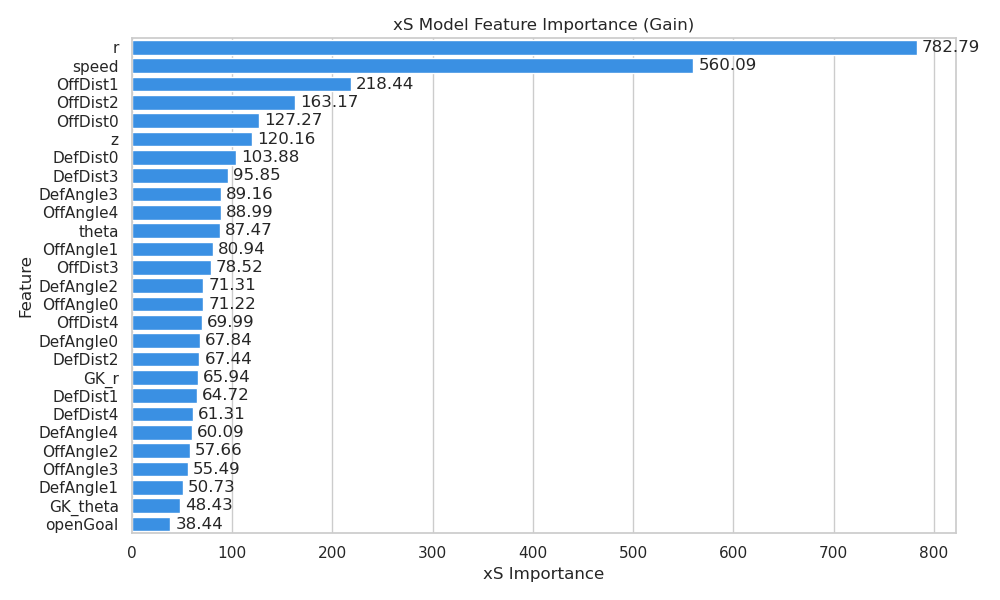}
  \caption{xS feature importance based on information gain}
  \label{xS_imp}
\end{figure}

\begin{figure}[H]
  \centering
  \includegraphics[width=0.9\textwidth]{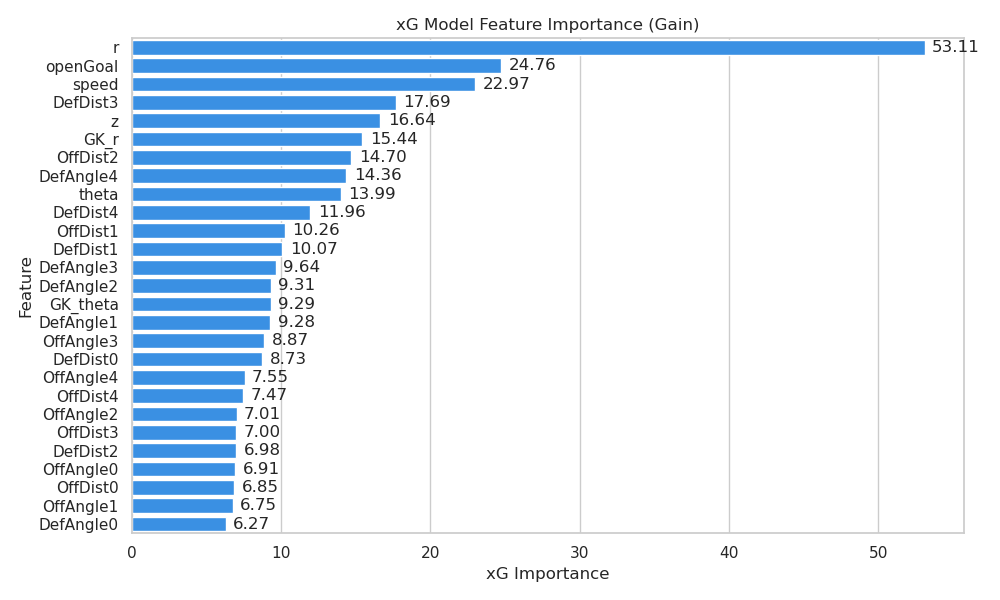}
  \caption{xG feature importance based on information gain}
  \label{xG_imp}
\end{figure}

It is useful to understand which features are most important for predicting the probabilities of taking (xS) and converting (xG) a shot. We plot feature importance by information gain for both models in Figures \ref{xS_imp} and \ref{xG_imp}. As expected, the distance from the ball to the goal (\texttt{r}) is the dominant predictor in both models. Additionally, ball speed ranks second for xS and third for xG, while the unobstructed goalmouth percentage (\texttt{openGoal}) is the second most influential feature for xG but contributes little to xS. This pattern is consistent with the idea that obstruction affects finishing quality more than the decision to shoot.

\begin{figure}[H]
  \centering
  \begin{subfigure}[b]{0.45\textwidth}
    \centering
    \includegraphics[width=\textwidth]{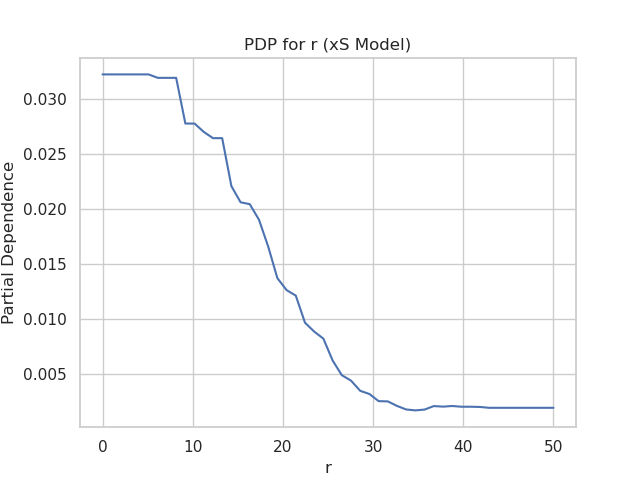}
    \subcaption{Distance from ball to goal}
    \label{xS_r}
  \end{subfigure}
  \hfill
  \begin{subfigure}[b]{0.45\textwidth}
    \centering
    \includegraphics[width=\textwidth]{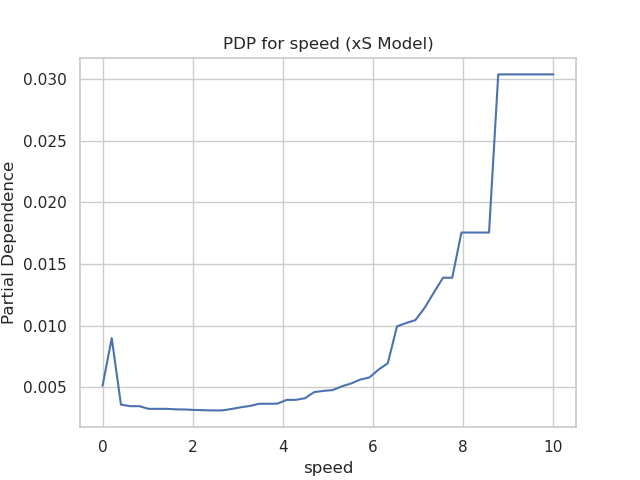}
    \subcaption{Ball speed}
    \label{xS_speed}
  \end{subfigure}
  \caption{Partial dependence plots (PDPs) for key variables affecting xS.}
  \label{fig:xS_PDPs}
\end{figure}

\begin{figure}[H]
  \centering
  \label{fig:xG_pdp}
  \begin{subfigure}[b]{0.45\textwidth}
    \centering
    \includegraphics[width=\textwidth]{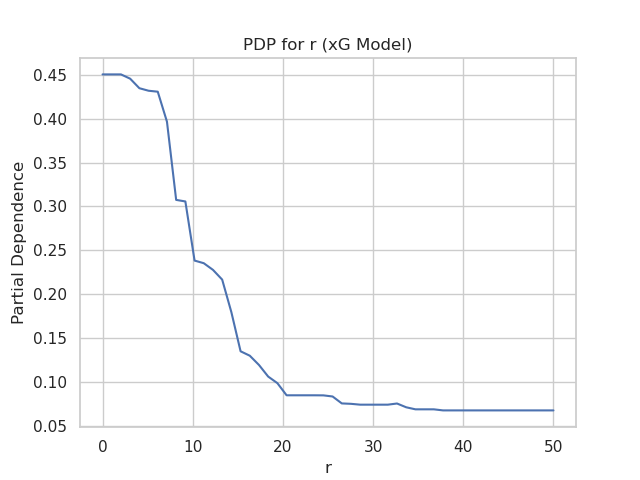}
    \subcaption{Distance from ball to goal}
    \label{xG_r}
  \end{subfigure}
  \hfill
  \begin{subfigure}[b]{0.45\textwidth}
    \centering
    \includegraphics[width=\textwidth]{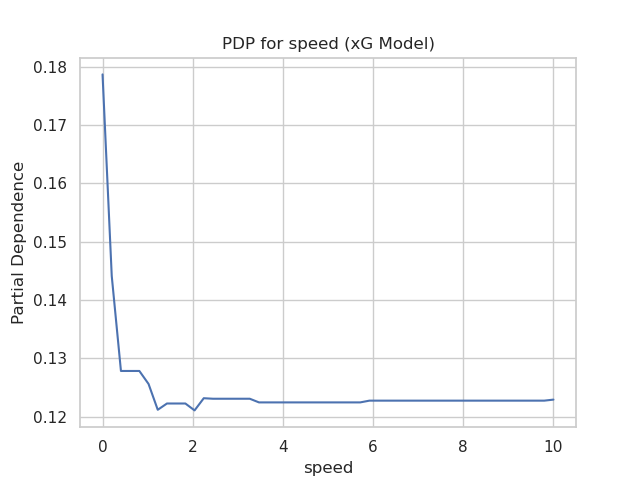}
    \subcaption{Ball speed}
    \label{xG_speed}
  \end{subfigure}

  \vspace{1em}

  \begin{subfigure}[b]{0.45\textwidth}
    \centering
    \includegraphics[width=\textwidth]{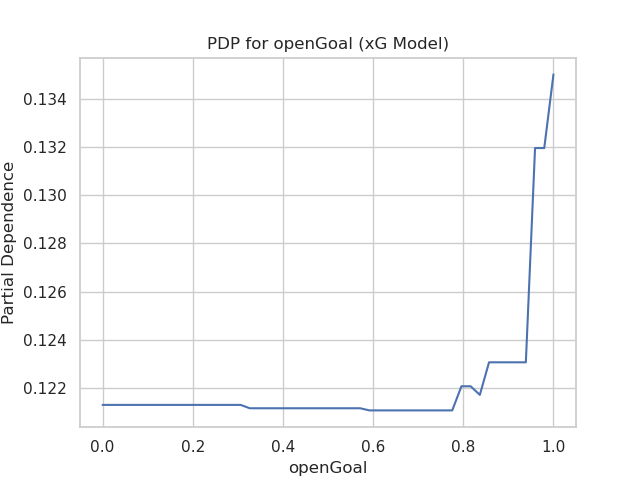}
    \subcaption{\texttt{openGoal}}
    \label{xG_openGoal}
  \end{subfigure}
  \hfill
  \begin{subfigure}[b]{0.45\textwidth}
    \centering
    \includegraphics[width=\textwidth]{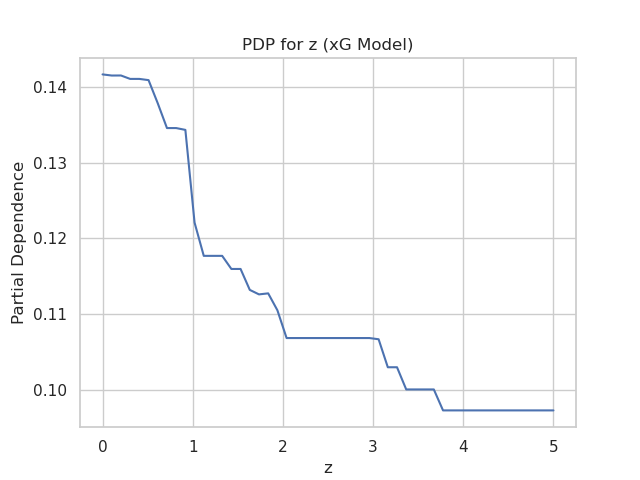}
    \subcaption{Ball height}
    \label{xG_z}
  \end{subfigure}

  \caption{Partial dependence plots (PDPs) for key variables affecting xG.}
  \label{fig:xG_PDPs}
\end{figure}

Additionally, the partial dependence plots in Figures \ref{fig:xS_PDPs} and \ref{fig:xG_PDPs} demonstrate the partial relationships between key features and predicted shooting (xS) and scoring (xG) probabilities. From these plots, it seems that both xS and xG decrease monotonically as the distance from the goal increases: from 8-30m for xS and 8-20m for xG. Ball speed exhibits opposing associations across tasks: higher speeds are linked to a greater likelihood of a shot in the next second (xS) but to lower scoring probability conditional on shooting (xG), with the latter dropping off sharply from an idle ball. A plausible interpretation is that fast sequences (e.g., through balls or crosses) often create chances, but from less controlled setups. For xG, \texttt{openGoal} is positively associated with conversion, whereas ball height is negatively associated. Ball height can also be considered a correlated proxy for what body part was used. These patterns align with soccer domain knowledge and suggest the models capture salient aspects of shot creation and conversion.


\subsection{Cross Validation Study on Goal Prediction}

To evaluate xG+ and xS against xG as predictive metrics, we use a rolling-origin cross-validation over three full EPL seasons (114 matchdays). We create one xG+, xS, and xG for each possession in a game and sum up those values across all possessions. The number in each possession is aggregated by one of these two approaches. 

\begin{enumerate}
  \item \textbf{At-least-one per possession:} $1 - \prod(1 - \text{xG+}_t)$
  \item \textbf{Max xG+ per possession:} $\max(\text{xG+}_t)$
\end{enumerate}

For xG, we also included a naive ``independent sum of all shots'' to match what is currently done with xG: aggregating values across an entire match.

 A full list of metrics is shown in Table \ref{tab:all_metrics_modeled}.

\begin{table}[H]
\centering
\caption{Metrics adjusted to model future team scoring}
\label{tab:all_metrics_modeled}
\begin{tabular}{lll}
\toprule
 & \textbf{Metric} & \textbf{Aggregation method} \\
\midrule
1. & xG   & At-least-one-per-possession \\
2. & xG   & Max-per-possession \\
3. & xG   & Sum-of-shots \\
4. & xS   & At-least-one-per-possession \\
5. & xS   & Max-per-possession \\
6. & xG+  & At-least-one-per-possession \\
7. & xG+  & Max-per-possession \\
\bottomrule
\end{tabular}
\end{table}

For each metric listed in Table \ref{tab:all_metrics_modeled} we perform a cross-validation study to predict team scoring on out-of-sample matches. We treat each matchday as a fold, train on the remaining 113 matchdays, and evaluate performance on the held-out one.

Within each training set, we fit a mixed-effects Poisson model to the chosen metric (Table \ref{tab:all_metrics_modeled}) with random intercepts for season, the attacking team, and the defending team, along with a fixed effect for home advantage. We specify our model in the \texttt{lme4} package in the R programming language as follows:
\begin{equation*}
\texttt{metric} \sim (1|\texttt{season}) + (1|\texttt{season:team}) + (1|\texttt{season:opp}) + \texttt{home}
\end{equation*}

This specification yields season-specific estimates of (i) team attacking strength, (ii) opponent defensive strength, (iii) season-level variation, and (iv) home advantage for each metric. These adjusted team- and opponent-level metrics are then fed into a second-stage Poisson model for goals, which we fit in \texttt{R} with the following formula:
\begin{equation*}
\texttt{goals} \sim \texttt{home} + \texttt{season} + \texttt{team\_off} + \texttt{opp\_def}
\end{equation*}

This two-stage modeling approach mirrors common practices in sports analytics, where underlying player and team skill estimates are used to forecast observable outcomes. We display our average cross-validation errors for each metric and aggregation strategy in Tables \ref{tab:mse} and \ref{tab:mae}.

\begin{table}[H]
\centering
\caption{Mean squared error (MSE) by metric and aggregation method}
\label{tab:mse}
\begin{tabular}{lrrr}
\toprule
\textbf{Aggregation method} & \textbf{xG+} & \textbf{xS} & \textbf{xG} \\
\midrule
At-least-one-per-possession & 2.84 & 2.90 & 2.94 \\
Max-per-possession          & 2.84 & 2.87 & 2.91 \\
Sum-of-shots                & ---  & ---  & 2.90 \\
\bottomrule
\end{tabular}
\end{table}

\begin{table}[H]
\centering
\caption{Mean absolute error (MAE) by metric and aggregation method}
\label{tab:mae}
\begin{tabular}{lrrr}
\toprule
\textbf{Aggregation method} & \textbf{xG+} & \textbf{xS} & \textbf{xG} \\
\midrule
At-least-one-per-possession & 1.86 & 1.87 & 1.89 \\
Max-per-possession          & 1.86 & 1.86 & 1.89 \\
Sum-of-shots                & ---  & ---  & 1.87 \\
\bottomrule
\end{tabular}
\end{table}

Across all specifications, xG+ yielded the lowest error, suggesting that combining shot probability (xS) with goal probability given a shot (xG) improves team-level forecasts relative to either component alone. The fact that xS also outperforms xG implies that short-horizon shot creation is a more stable team-level signal than shot quality, underscoring the value of modeling latent chances and the sequential structure of possessions.

To complement the average errors in Tables \ref{tab:mse} and \ref{tab:mae}, we also examine the variability in performance of each specification across matchdays. For each metric and aggregation method, we compute the empirical 90\% interval of squared error over cross-validation folds. These intervals are reported in Table \ref{tab:sq_err_intervals}.

\begin{table}[ht]
\centering
\caption{90\% squared error intervals by metric and aggregation method}
\label{tab:sq_err_intervals}
\begin{tabular}{lccc}
\toprule
\textbf{Aggregation method} & \textbf{xG+} & \textbf{xS} & \textbf{xG} \\
\midrule
At-least-one-per-possession & (0.800, 2.25) & (0.823, 2.28) & (0.819, 2.32) \\
Max-per-possession          & (0.792, 2.26) & (0.817, 2.27) & (0.802, 2.31) \\
Sum-of-shots                & ---           & ---           & (0.822, 2.30) \\
\bottomrule
\end{tabular}
\end{table}

Across Tables \ref{tab:mse}, \ref{tab:mae}, and \ref{tab:sq_err_intervals}, the max-per-possession aggregation is either indistinguishable (to two decimal places) from, or slightly better than, the at-least-one-per-possession approach. While the at-least-one rule is more directly interpretable as a cumulative scoring probability over the course of a possession, it also grows mechanically with possession length. By contrast, the max operator focuses on the single most dangerous moment in the possession and appears to deliver equal or better predictive accuracy for all three metrics.

Although xG+ achieves the lowest mean error in Tables \ref{tab:mse} and \ref{tab:mae}, the 90\% intervals in Table \ref{tab:sq_err_intervals} still overlap across methods. This is not surprising: some matchdays feature unusually volatile or upset-heavy results, which limit the separation we can observe in aggregate error summaries. To probe this further, we examine performance on a matchday-by-matchday basis. For each aggregation method, we compute the fraction of matchdays on which its mean squared error is lower than that of the traditional sum-of-xG benchmark (Table \ref{tab:fraction_better_than_xGsum}). If two approaches are truly equivalent, we would expect each to outperform the baseline on roughly 50\% of matchdays.

\begin{table}[ht]
\centering
\caption{Number and fraction of matchdays on which each specification outperforms the sum of xG}
\label{tab:fraction_better_than_xGsum}
\begin{tabular}{lccc}
\toprule
\textbf{Aggregation method} & \textbf{xG+} & \textbf{xS} & \textbf{xG} \\
\midrule
At-least-one-per-possession & 69 (0.605) & 56 (0.491) & 50 (0.439) \\
Max-per-possession          & 70 (0.614) & 60 (0.526) & 56 (0.491) \\
\bottomrule
\end{tabular}
\end{table}

Under the null hypothesis that each alternative method is equally likely to beat the sum-of-xG baseline on any given matchday, the number of wins over 114 matchdays follows a $\text{Binomial}(114, 0.5)$ distribution, assuming performance across matchdays is independent. Observing 69 or more wins for the at-least-one xG+ specification has probability 0.015, and observing 70 or more wins for the max-per-possession xG+ specification has probability 0.009. Taken together, these results provide strong evidence that xG+ offers a genuinely better predictor of future team scoring than the traditional xG sum-of-shots approach, and that its apparent gains are unlikely to be explained by sampling variability alone. We validate this claim further with a training sample robustness analysis included in Appendix \ref{app:b}.

\subsection{Player Evaluation}

Traditional xG-based metrics often suggest that few players consistently ``outperform'' their expected goals over multiple seasons \citep{goodman2014finishing,statsbomb2017finishing,davis2024biases}. In other words, the extent to which a player scores above or below the sum of their xG in one season is only weakly informative about whether they will over- or under-perform in the future. 

To formalize this, we define three player-season performance-over-expected quantities:
\[
\text{GOE}_{xG} = \text{Goals} - \text{xG}, \qquad
\text{SOE} = \text{Shots} - \text{xS}, \qquad
\text{GOE}_{xG+} = \text{Goals} - \text{xG+}.
\]
For xG and xG+, performance over expected is defined relative to goals; for xS it is defined relative to shots. Table~\ref{tab:metricStability} reports the year-to-year correlations of these measures at the player-season level. Consistent with prior work, goals over expected relative to xG exhibit very low stability (correlation $\approx 0.12$). By contrast, shots over expected relative to xS are highly persistent across seasons (correlation $\approx 0.63$), while performance over xG+ lies between these extremes, reflecting the fact that xG+ combines information from both shot creation and finishing.

\begin{table}[!h]
\centering
\caption{Year-to-year correlation (stability) of performance over expected}
\label{tab:metricStability}
\begin{tabular}[t]{rrr}
\toprule
\textbf{GOE\(_{xG}\)} & \textbf{SOE} & \textbf{GOE\(_{xG+}\)} \\
\midrule
0.12 & 0.63 & 0.35 \\
\bottomrule
\end{tabular}
\end{table}

This contrast aligns with the motivating example in Figure~\ref{fig:mbappeShotCreate}, featuring Kylian Mbappé. Traditional xG assigned his eventual shot a high value (0.5 at FBRef), a level that would be difficult to consistently outperform in finishing alone. However, that valuation ignores the skill required to create such a high-probability opportunity in the first place. By jointly modeling the probability of creating a shot (xS) and converting it (xG), our framework attributes value to both the buildup and the finish, ensuring that Mbappé and similar players receive credit for repeatedly manufacturing high-quality chances, not just for converting them.

To study this effect at scale, we focus first on \emph{shots over expected} (SOE), as defined above, and examine which players most strongly over- or under-perform this baseline. Table~\ref{tab:xsoe-summary} reports the top 10 and bottom 10 player-seasons by SOE in our dataset (EPL 2022--2025). Players with large positive SOE are consistently able to turn possession states into more shots than expected, while those with large negative SOE tend to generate fewer shots than their xS would predict.

\begin{table}[h]
\centering
\caption{Top 10 and bottom 10 player-seasons by shots over expected (SOE)}
\label{tab:xsoe-summary}
\begin{tabular}{l c l r r r r}
\toprule
 & \textbf{Season} & \textbf{Player name} & \textbf{Shots} & \textbf{xS} & \textbf{SOE} & \textbf{Matches} \\
\midrule
 & 2022-23 & Marcus Rashford   & 103 & 54.4 & 48.6  & 35 \\
 & 2023-24 & Mohamed Salah     & 114 & 67.0 & 47.0  & 32 \\
 & 2024-25 & Antoine Semenyo   & 105 & 60.0 & 45.0  & 36 \\
 & 2023-24 & Erling Haaland    & 112 & 67.5 & 44.5  & 31 \\
 & 2023-24 & Darwin Núñez      & 98  & 55.2 & 42.8  & 35 \\
 & 2024-25 & Cole Palmer       & 114 & 71.8 & 42.2  & 37 \\
 & 2022-23 & Harry Kane        & 116 & 74.3 & 41.7  & 38 \\
 & 2022-23 & Erling Haaland    & 120 & 78.4 & 41.6  & 35 \\
 & 2024-25 & Eberechi Eze      & 92  & 50.9 & 41.1  & 34 \\
 & 2024-25 & Matheus Cunha     & 92  & 51.2 & 40.8  & 33 \\
\midrule
 & 2024-25 & Rico Lewis        & 11  & 22.7 & -11.7 & 28 \\
 & 2024-25 & Joško Gvardiol    & 35  & 46.9 & -11.9 & 37 \\
 & 2024-25 & Thomas Partey     & 26  & 38.2 & -12.2 & 35 \\
 & 2023-24 & Manuel Akanji     & 9   & 21.3 & -12.3 & 29 \\
 & 2022-23 & Ben White         & 9   & 23.2 & -14.2 & 38 \\
 & 2022-23 & Pascal Gro\ss     & 36  & 50.8 & -14.8 & 37 \\
 & 2024-25 & İlkay Gündogan    & 28  & 42.8 & -14.8 & 31 \\
 & 2023-24 & Bernardo Silva    & 30  & 45.0 & -15.0 & 33 \\
 & 2022-23 & Bruno Guimarães   & 22  & 37.1 & -15.1 & 32 \\
 & 2024-25 & Bernardo Silva    & 27  & 43.1 & -16.1 & 33 \\
\bottomrule
\end{tabular}
\end{table}

The pattern in Table~\ref{tab:xsoe-summary} is consistent with the stability results in Table~\ref{tab:metricStability}: players with large positive SOE are, in general, widely regarded as among the Premier League's most impactful attackers. Their primary repeatable skill is not persistently finishing above xG, but rather consistently creating more shots than we would expect given the state of play.

\begin{figure}[ht!]
\centering
\includegraphics[width=4.5in]{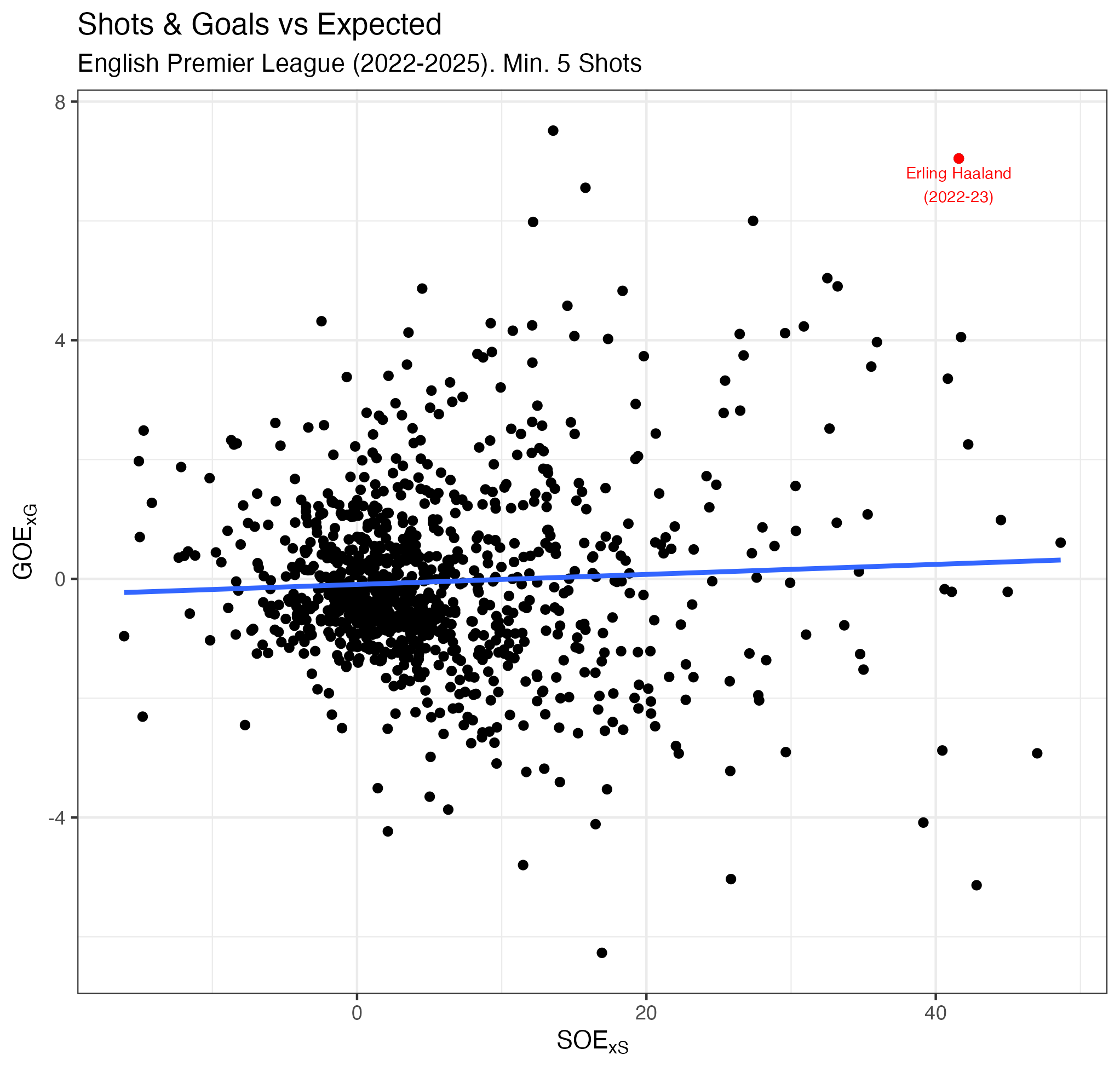}
\caption{Goals over expected relative to xG ($\text{GOE}_{xG}$) vs shots over expected (SOE) for EPL player-seasons, 2022--2025.}
\label{fig:xseoe_vs_xgoe}
\end{figure}

The joint behavior of shot creation and finishing is illustrated in Figure~\ref{fig:xseoe_vs_xgoe}, which plots shots over expected (SOE) against goals over expected relative to xG ($\text{GOE}_{xG}$). Erling Haaland's record-breaking 2022--23 season, in which he scored 36 Premier League goals, appears in the upper-right region of this plot: he not only generates more shots than expected (high SOE), but also benefits from substantially favorable finishing variance relative to his xG (high $\text{GOE}_{xG}$).

Finally, we compare players by their over-performance relative to xG+ and xG on a per-match basis. For each player-season, we scale the performance-over-expected quantities by matches played (MP):
\[
\text{GOE}^{\text{pm}}_{xG+} = \frac{\text{GOE}_{xG+}}{\text{MP}}
\quad\text{and}\quad
\text{GOE}^{\text{pm}}_{xG} = \frac{\text{GOE}_{xG}}{\text{MP}}.
\]
Table~\ref{tab:xSvsXG} lists the top ten player-seasons by $\text{GOE}^{\text{pm}}_{xG+}$ and $\text{GOE}^{\text{pm}}_{xG}$, respectively, among players with at least 10 matches with a chance. The xG+ ranking places sustained shot creation and chance generation at the top of the list (e.g., Haaland, Isak, Kane, Salah), aligning more closely with long-run attacking impact, whereas the xG-only ranking is more sensitive to shorter-run finishing streaks.

\begin{table}[ht!]
\centering\footnotesize
\setlength{\tabcolsep}{2pt}
\renewcommand{\arraystretch}{0.9}
\caption{Top 10 player-seasons ranked by $\text{GOE}^{\text{pm}}_{xG+}$ and $\text{GOE}^{\text{pm}}_{xG}$, EPL 2022--2025}
\label{tab:xSvsXG}

\begin{minipage}{0.45\textwidth}
\centering
\textbf{Goals over expected relative to xG+}\\[2pt]
\begin{tabular}{llccc}
\toprule
 & \textbf{Player} & $\text{GOE}^{\text{pm}}_{xG+}$ & \textbf{Goals} & \textbf{Matches} \\
\midrule
22--23 & Erling Haaland     & 0.52 & 36 & 35 \\
23--24 & Erling Haaland     & 0.44 & 30 & 31 \\
24--25 & Omar Marmoush      & 0.41 & 10 & 16 \\
24--25 & Yoane Wissa        & 0.35 & 22 & 35 \\
24--25 & Mohamed Salah      & 0.33 & 27 & 38 \\
24--25 & Chris Wood         & 0.31 & 23 & 36 \\
23--24 & Cole Palmer        & 0.30 & 19 & 34 \\
24--25 & Alexander Isak     & 0.30 & 25 & 34 \\
23--24 & Alexander Isak     & 0.30 & 19 & 28 \\
22--23 & Harry Kane         & 0.29 & 25 & 38 \\
\bottomrule
\end{tabular}
\end{minipage}
\hspace{0.02\textwidth}
\begin{minipage}{0.45\textwidth}
\centering
\textbf{Goals over expected relative to xG}\\[2pt]
\begin{tabular}{llccc}
\toprule
 & \textbf{Player} & $\text{GOE}^{\text{pm}}_{xG}$ & \textbf{Goals} & \textbf{Matches} \\
\midrule
24--25 & Omar Marmoush      & 0.41 & 10 & 16 \\
24--25 & Chris Wood         & 0.31 & 23 & 36 \\
24--25 & Michael Keane      & 0.21 &  3 & 10 \\
22--23 & Erling Haaland     & 0.52 & 36 & 35 \\
22--23 & Roberto Firmino    & 0.25 & 10 & 21 \\
22--23 & Martin \"{O}degaard & 0.22 & 15 & 37 \\
23--24 & Heung-min Son      & 0.28 & 20 & 35 \\
22--23 & Matias Vi\~{n}a    & 0.26 &  3 & 10 \\
22--23 & Alexander Isak     & 0.24 & 11 & 22 \\
23--24 & Taiwo Awoniyi      & 0.26 &  7 & 19 \\
\bottomrule
\end{tabular}
\end{minipage}

\vspace{0.2cm}
\footnotesize\emph{Note: Sample restricted to player-seasons with at least 10 matches with a shot opportunity.}
\end{table}

\newpage

\section{Discussion}

\subsection{Conclusions}

Our proposed xG+ metric addresses key limitations of existing expected goals models by explicitly modeling the probability of a shot and incorporating this into a possession-based framework, which allows us to account for rebounded chances, credit dangerous non-shot moments, and more accurately evaluate team and player strength.

Our analysis shows that xG+ is more predictive of actual goals than traditional methods like xG, and that shot creation ability is far more stable across seasons than finishing ability. These insights can inform recruitment, tactical analysis, and performance forecasting.

\subsection{Potential Limitations}

Our study has several limitations that qualify the interpretation of our findings. First, measurement noise in video tracking can introduce error in ball and player locations which our models depend on. Furthermore, our models -- trained exclusively on the 2022–2025 EPL seasons -- may not generalize well to other leagues or competitions without recalibration. Additionally, the one-second horizon used to define xS also renders labels sensitive to small timestamp misalignments, creating boundary effects near the decision window. On the data-curation side, our requirement for “clear possession in the attacking third” depends on noisy possession indicators and may inadvertently exclude genuine goal threats (e.g., dangerous crosses or through balls with no attackers nearby). Feature construction introduces further approximation: \texttt{openGoal} treats players as identical 2D occluders and ignores differences in reach, height, and jumping. Methodologically, the framework scores frames with snapshot features and thus omits sequential dependence -- such as recovery runs, second balls, and pass–shoot chains -- that may shape both shot taking and shot quality. At the aggregation stage, possession-level summaries of xS can mechanically reward longer possessions even after the optimal shooting moment has passed, so caution is warranted when applying the metric to player evaluation. Finally, selection on opportunity persists: stronger teams and players reach dangerous states more frequently, and this exposure is not explicitly modeled in the present formulation.

\subsection{Future Work}

There are several natural extensions to this work. Methodologically, replacing snapshot models with sequence models -- such as temporal point processes or survival/hazard formulations -- may provide richer estimation of xS. Additionally, higher-fidelity tracking may yield improved feature construction (e.g., exact player and ball positions, player orientation), yielding more reliable estimates. Adding hierarchical team and player effects would also mitigate selection on opportunity. Decision-analytic extensions include estimating the counterfactual value of actions (e.g., shooting now vs.\ continuing the possession), which may enable policy-aware variants of xG+. On the defensive side, this framework could be mirrored to quantify shot and goal suppression, with credit assignment for lane-closing and goalkeeper positioning. Finally, external validation across other leagues and competitions, together with real-time implementations of \texttt{openGoal}, xS, and xG+ would broaden applicability for scouting, broadcasting, and in-match decision-making.

\bibliographystyle{apalike}
\bibliography{references}

\newpage

\appendix

\section{Description of Model Features}\label{app:a}

\begin{figure}[H]
\centering
  \includegraphics[width=0.9\textwidth]{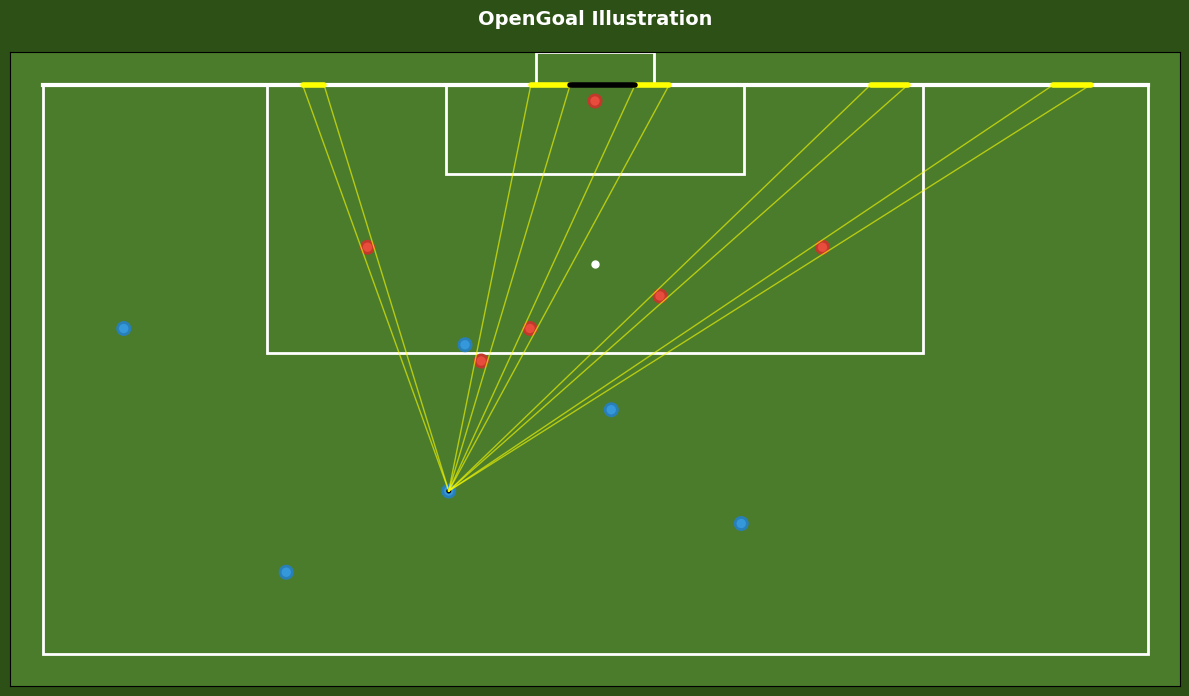}
  \caption{An illustration of how \texttt{openGoal} is constructed. Red dots correspond to defenders, yellow lines represent tangent line pairs, yellow segments represent obstructed portions of the goal line, and the black segment represents the "open" portion of the goalmouth.}
  \label{fig:openGoal}
\end{figure}

Table \ref{tab:data-dictionary} includes a full description of the features used to train the xS and xG models in Section \ref{sec:model-estimation}. We compute \texttt{openGoal} by modeling defenders (excluding the goalkeeper) as uniform circles with 75cm diameters and computing the tangent line pairs from the ball to every defender between the ball and goal. The segments made up by the intersections between tangent line pairs and the goal line are considered ``obstructed.'' \texttt{openGoal} is computed as the percentage of the goal not covered by the obstructed segments. An illustration of this is presented in Figure \ref{fig:openGoal}; red dots correspond to defenders, yellow lines represent tangent line pairs, yellow segments represent obstructed portions of the goal line, and the black segment represents the "open" portion of the goalmouth.

\begin{table}[htbp]
\centering
\small
\setlength{\tabcolsep}{6pt}
\renewcommand{\arraystretch}{1.1}
\caption{Data dictionary of features used to train xS and xG models}
\label{tab:data-dictionary}
\begin{tabular}{@{}p{0.28\textwidth}@{}p{0.16\textwidth}@{}p{0.56\textwidth}@{}}
\toprule
\textbf{Variable} & \textbf{Units} & \textbf{Description} \\
\midrule

\multicolumn{3}{@{}l}{\textit{Ball features}}\\
\texttt{r}      & m   & Distance from ball to center of goal. \\
\texttt{theta}  & rad & Angle from ball to center of goal. \\
\texttt{z}      & m   & Ball height above pitch. \\
\texttt{speed}  & m/s & Ball speed. \\

\addlinespace[4pt]
\multicolumn{3}{@{}l}{\textit{Goal features}}\\
\texttt{openGoal} & [0,1] & Unobstructed share of the goal mouth; see Fig.~\ref{fig:openGoal}. \\

\addlinespace[4pt]
\multicolumn{3}{@{}l}{\textit{Goalkeeper features}}\\
\texttt{GK\_r}      & m   & Distance from goalkeeper to center of goal. \\
\texttt{GK\_\!theta} & rad & Bearing from goalkeeper to center of goal. \\

\addlinespace[4pt]
\multicolumn{3}{@{}l}{\textit{Outfield player features}}\\
\texttt{DefAngle\_\{0..4\}} & rad & Bearing from ball to the $k+1$-th nearest defender (non-GK), where $k\in\{0,\dots,4\}$. \\
\texttt{DefDist\_\{0..4\}}  & m   & Distance from ball to the $k+1$-th nearest defender (non-GK), where $k\in\{0,\dots,4\}$. \\
\texttt{OffAngle\_\{0..4\}} & rad & Bearing from ball to the $k+1$-th nearest attacker (excluding carrier), $k\in\{0,\dots,4\}$. \\
\texttt{OffDist\_\{0..4\}}  & m   & Distance from ball to the $k+1$-th nearest attacker (excluding carrier), $k\in\{0,\dots,4\}$. \\
\bottomrule
\end{tabular}
\end{table}

\section{Analysis of Downstream Variance}\label{app:b}

Because our modeling pipeline is multi-stage, estimation error in the xS and xG components can propagate to downstream models. To assess how sensitive our conclusions are to the particular training sample, we perform a training-sample robustness check based on repeated subsampling at the matchday level.

We first designate 20\% of matchdays as a fixed hold-out test set, which is excluded from all model fitting. On the remaining 80\% of matchdays, we construct ten distinct training samples by randomly selecting 90\% of available matchdays (without replacement) for each replication. For each of these ten training samples, we repeat the full modeling pipeline from Section~\ref{sec:model-estimation} and the subsequent mixed-effects and goals models, and then generate goal predictions for all clubs in all matches in the fixed test set for each metric and aggregation strategy.

Table~\ref{tab:mse_interval_comparison} summarizes these results. For each metric and aggregation rule, we report the minimum and maximum difference in average mean squared error (MSE) relative to the traditional xG game-sum benchmark across the ten replications. Negative values indicate that a given method improves upon the sum-of-xG baseline on the held-out test matches.

For both aggregation strategies, all three possession-level specifications (xG+, xS, and xG) exhibit uniformly negative differences, indicating lower test-set MSE than the traditional xG game-sum in every training subsample. The additional sampling variability introduced by this resampling scheme makes it difficult to cleanly rank the three alternatives against one another, but it does clearly show that each possession-based approach improves on the traditional independent sum-of-shots xG benchmark. Taken together with the main-sample results, these findings suggest that the gains from xG+ are robust to variation in the training sample and are unlikely to be an artifact of a single favorable split of the data.

\begin{table}[ht]
\centering
\caption{Min--max intervals for the difference in average mean squared error (MSE) relative to the sum of xG}
\label{tab:mse_interval_comparison}
\begin{tabular}{lccc}
\hline
\textbf{Aggregation method} & \textbf{xG+ diff} & \textbf{xS diff} & \textbf{xG diff} \\
\hline
At-least-one-per-possession 
  & (\textbf{-0.112}, \textbf{-0.0612})
  & (\textbf{-0.122}, \textbf{-0.0722})
  & (\textbf{-0.146}, \textbf{-0.0957}) \\
Max-per-possession          
  & (\textbf{-0.0929}, \textbf{-0.0458})
  & (\textbf{-0.108}, \textbf{-0.0572})
  & (\textbf{-0.106}, \textbf{-0.0503}) \\
\hline
\end{tabular}
\end{table}

\end{document}